\documentclass{appolb}
\usepackage{epsfig}
\usepackage{amssymb}

\begin{document}

\title{Across the deconfinement\thanks{Presented at CPOD 2016, Wroclaw 30 May -
4 June 2016.} }

\author{C.~Bonati\thanks{Speaker.}, M.~D'Elia, M.~Mariti, M.~Mesiti, F.~Negro
\address{Dipartimento di Fisica dell'Universit\`a di Pisa and INFN - Sezione di
Pisa,\\ Largo Pontecorvo 3, I-56127 Pisa, Italy}
\and
F. Sanfilippo
\address{School of Physics and Astronomy, University of Southampton, \\ SO17 1BJ
Southampton, United Kingdom}
}

\maketitle

\begin{abstract}
The deconfinement transition at vanishing chemical potential can be reliably
studied by lattice simulations and its general features are by now well known.
On the contrary, what happens at finite density is still largely unknown and we
will review the results obtained in the last year regarding the dependence, for
small density, of the (pseudo)critical temperature on the baryonic chemical
potential.
\end{abstract}

\PACS{12.38.Aw, 
12.38.Gc, 
12.38.Mh 
}

\section{Introduction}

One of the goals of Lattice Quantum ChromoDynamics (LQCD) is to provide a first
principles description of the QCD phase diagram and of the nonperturbative
behaviour of the strongly interacting thermal medium.  Beyond their purely
theoretical interest, such informations are valuable for a better understanding
of several phenomenologically relevant processes, among which heavy ions
ultrarelativistic collisions play a dominant role.

The study of the thermodynamical properties of QCD at vanishing baryon density
by means of LQCD numerical simulations is a standard (although not
computationally easy) task.  When a nonzero baryon density is present this is
no more true: the usual importance sampling Monte Carlo methods used in LQCD
simulations cannot be used anymore, due to the well known sign problem.

The only physically relevant case in which the sign problem can be circumvented
is the case of small baryon density or, more precisely, small baryon chemical
potential $\mu_B$: if we are interested in the observable $A(\mu_B)$, we can
expand it as $A(\mu_B)\simeq A_0+A_1\mu_B+A_2\mu_B^2+\cdots$ and try to
estimate the coefficients of the series. Which values of $\mu_B$ are ``small
enough'' for this procedure to give reliable results is something that can only
be determined \emph{a posteriori}, given the accuracy that one can obtain in
the evaluation of the expansion coefficients.  

This general idea can be implemented in two different ways: in the approach
known as the Taylor expansion method, the coefficient $A_n$ is obtained by
evaluating $\partial_{\mu_B}^nA(\mu_B)|_{\mu_B=0}$ using standard $\mu_B=0$
simulations \cite{Allton:2002zi}.  In the analytic continuation approach,
simulations performed at imaginary values of the chemical potential
$\mu_B=i\mu_{B,I}$ (at which no sign problem is present) are performed, and the
coefficients $A_n$ are then extracted by fitting the results \cite{deFPh, DEL}.
The two approaches have complementary advantages and drawbacks: in the Taylor
expansion method the coefficients $A_n$ are directly obtained, but their
estimators become more and more noisy as $n$ is increased; in the analytic
continuation method the $A_n$ values have to be fitted, but one can use
observables with good signal to noise ratios.
 
A property of the QCD phase diagram that can be investigated by using LQCD
simulations is the dependence, at least for small values of $\mu_B$, of the
(pseudo)critical temperature on the baryon chemical potential: $T_c(\mu_B)$ can
be developed in even powers of $\mu_B$, and the curvature $\kappa$ of the
critical line is defined by
\begin{equation}\label{curv}
T_c(\mu_B)/T_c=1-\kappa (\mu_B/T_c)^2\, +\, O(\mu_B^4/T_c^4)\ ,
\end{equation}
where $T_c$ denotes the (pseudo)critical temperature at $\mu_B=0$.  It is
interesting to compare the curvature of the QCD (pseudo)critical line with the
curvature of the freeze-out curve extracted from heavy ion collisions, see e.g.
\cite{Cleymans:2005xv, Becatt}. While there is no
compelling theoretical reason for the critical and the freeze-out lines to
coincide (in fact the first is an equilibrium property of QCD while the second
depends also on out-of-equilibrium properties of the strongly interacting
medium), a precise quantitative comparison of these observables could help in
better understanding the physical processes involved in the cooling of the
quark-gluon plasma.

\section{Numerical results}

\begin{figure}[t]
\centering
\includegraphics[width=0.55\textwidth,clip]{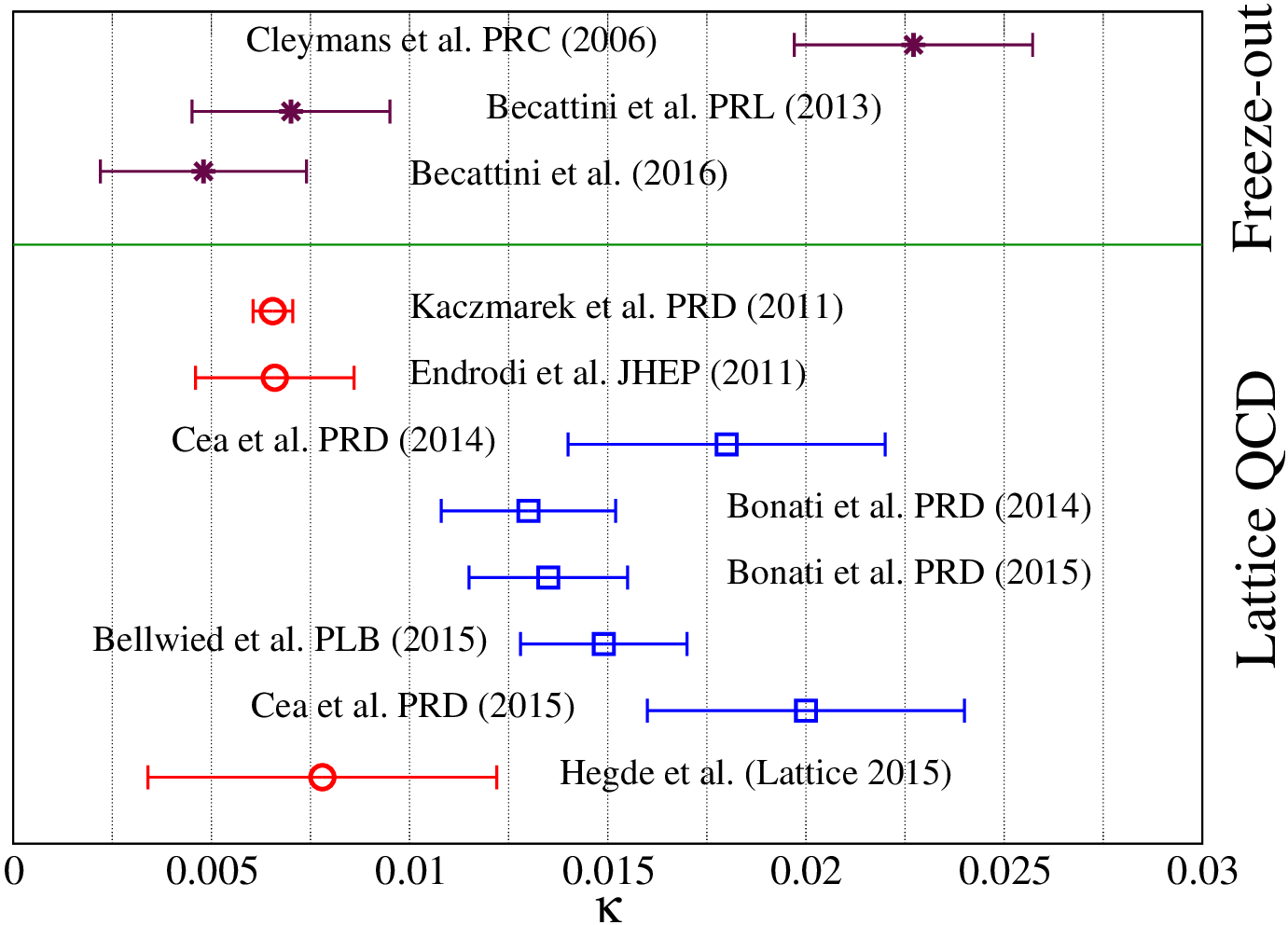}
\caption{Recent determinations of the curvature of the freeze-out curve (from
\cite{Cleymans:2005xv, Becatt}) and of the QCD
(pseudo)critical line (from \cite{Kaczmarek:2011zz, Endrodi:2011gv, Bari, Pisa,
Bellwied:2015rza, Hegde:2015tbn}). Red circles denote data obtained by the
Taylor expansion method, while blue squares correspond to data obtained by the
analytic continuation method.}
\label{fig:summary}
\end{figure}

At vanishing baryon chemical potential a real phase transition is not present,
but just an analytic crossover; as a consequence it is important to specify the
observable used in the determination of $T_c(\mu_B)$, since different
observables can in principle lead to different results. All LQCD results
that will be discussed in the following refer to observables related to the
restoration of the chiral symmetry (chiral condensate or chiral
susceptibility), see Fig.~\ref{fig:summary} for a graphical summary of the
results.

\begin{figure}[t]
\centering
\includegraphics[width=0.48\textwidth,clip]{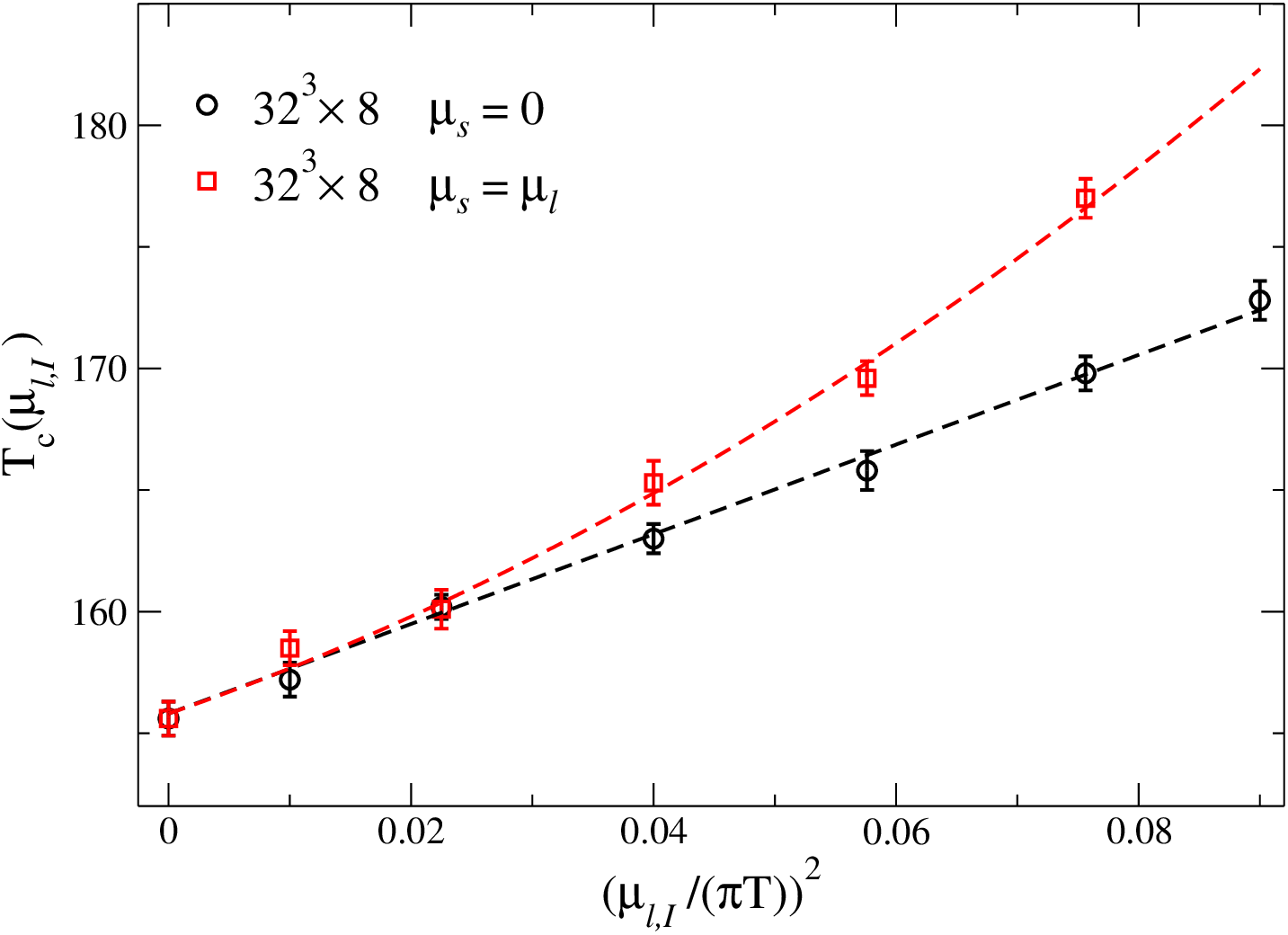}
\hspace{0.2cm}
\includegraphics[width=0.485\textwidth,clip]{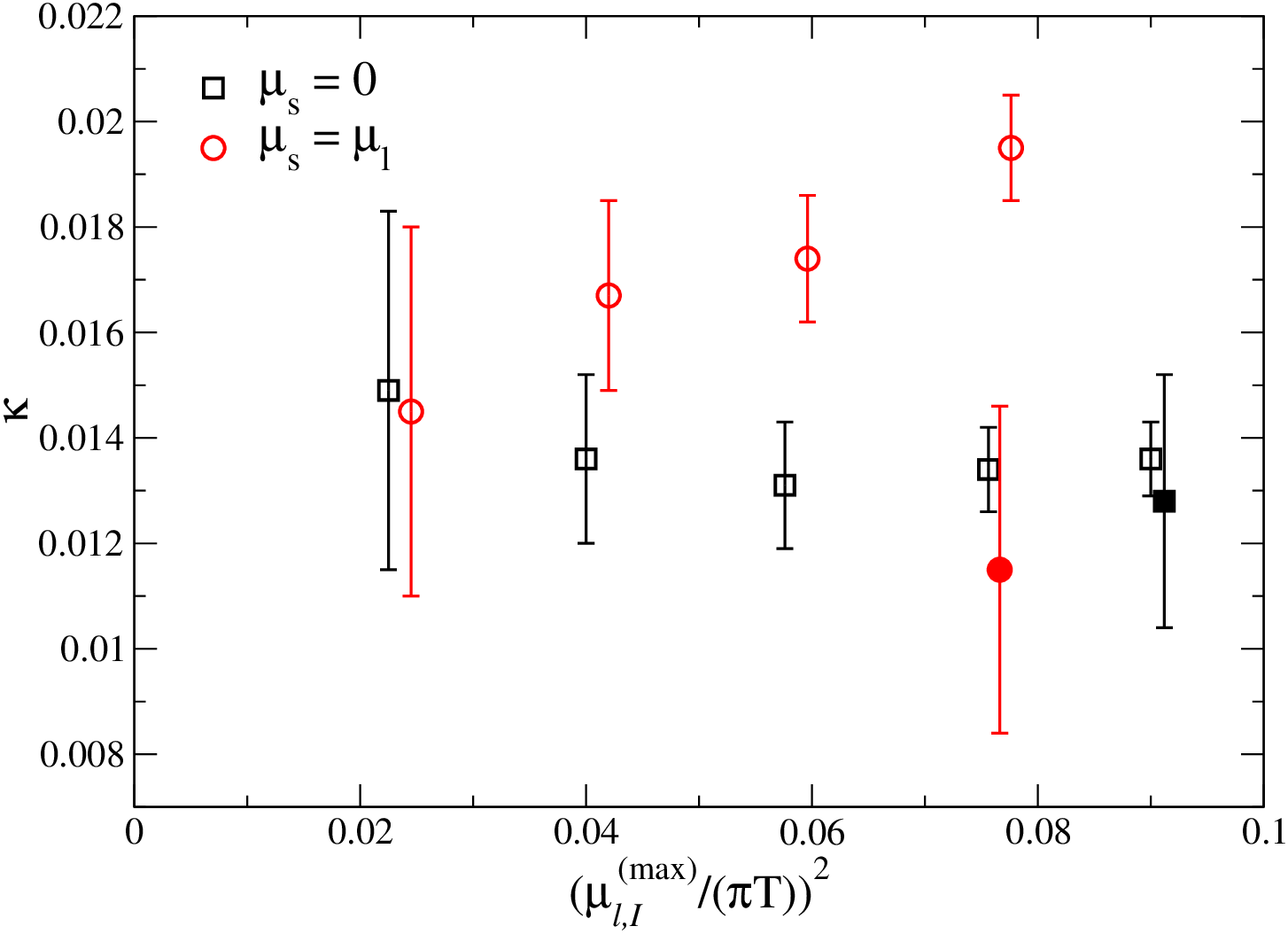}
\caption{(left) Dependence of the critical temperature on the (imaginary)
light chemical potential for the two setups $\mu_s=\mu_l$ and $\mu_s=0$.
(right) Results of the fit to extract $\kappa$ for different fit ranges; empty
symbols denote the purely quadratic fit, while also the quartic correction is
used for the filled symbols (from \cite{Pisa})}
\label{fig:syst}
\end{figure}

Given the great phenomenological significance of the curvature of the QCD
(pseudo)critical line, after the seminal works \cite{Kaczmarek:2011zz,
Endrodi:2011gv} several LQCD groups recently got involved in new computations
of $\kappa$, with the principal aim of improving the control of the
systematics.  The analytic continuation method is attractive in this respect,
since it enables to compute $T_c(\mu_B)$ using the same standard procedures
that are used in the $\mu_B=0$ case: looking for the inflection point of the
chiral condensate or the maximum of the chiral susceptibility
$\chi_{\bar{\psi}\psi}$ (see \cite{Pisa} for a discussion of the
effect on $\kappa$ of different renormalization procedures). A new source
of systematics is in this case the fitting procedure, that however can be kept
well under control by checking, e.g., the dependence of the results on the fit
range used, see Fig.~\ref{fig:syst}.

Another improvement with respect to the first determinations is a better
understanding of the role played by the quark chemical potentials: for the
results to be of direct physical relevance in heavy ions collisions, the
chemical potentials have to be tuned in such a way that the total strangeness
vanishes and the electric charge is related to the baryon number by $\langle
Q\rangle = r \langle B\rangle$, with $r\simeq 0.4$. The first studies
\cite{Kaczmarek:2011zz, Endrodi:2011gv} adopted the chemical potential setup
$\mu_u=\mu_d=\mu_B/3$, $\mu_s=0$ and in \cite{Bari} the setup
$\mu_u=\mu_d=\mu_s=\mu_B/3$ was used, while strangeness neutrality (near the
transition) implies $\mu_s\approx \mu_u/4$ (see \cite{chpot}). In \cite{Pisa}
it was shown that the value of $\kappa$ is insensitive (within the numerical
accuracy) to the value of $\mu_s$, while higher orders in the development of
$T_c(\mu_B)$ depends on $\mu_s$ (see Fig.~\ref{fig:syst}), a fact that can
explain the slightly larger values of $\kappa$ obtained in \cite{Bari} and is
likely related to the so called Roberge-Weiss transition, taking place at
$\mu_u=\mu_d=\mu_s=i\pi T/3$ \cite{Bonati:2016pwz}.  A further confirmation of
the $\mu_s$-independence of $\kappa$ is given by the results of
\cite{Bellwied:2015rza}: a value of $\kappa$ in very good agreement with the
ones of \cite{Pisa} is obtained by performing simulations directly at
strangeness neutrality.  The effect on $\kappa$ of the isospin breaking
constraint $\langle Q\rangle =0.4\langle B\rangle$ was also shown in
\cite{Bellwied:2015rza} to be negligible with the present accuracy.

\begin{figure}[b]
\centering
\includegraphics[width=0.45\textwidth,clip]{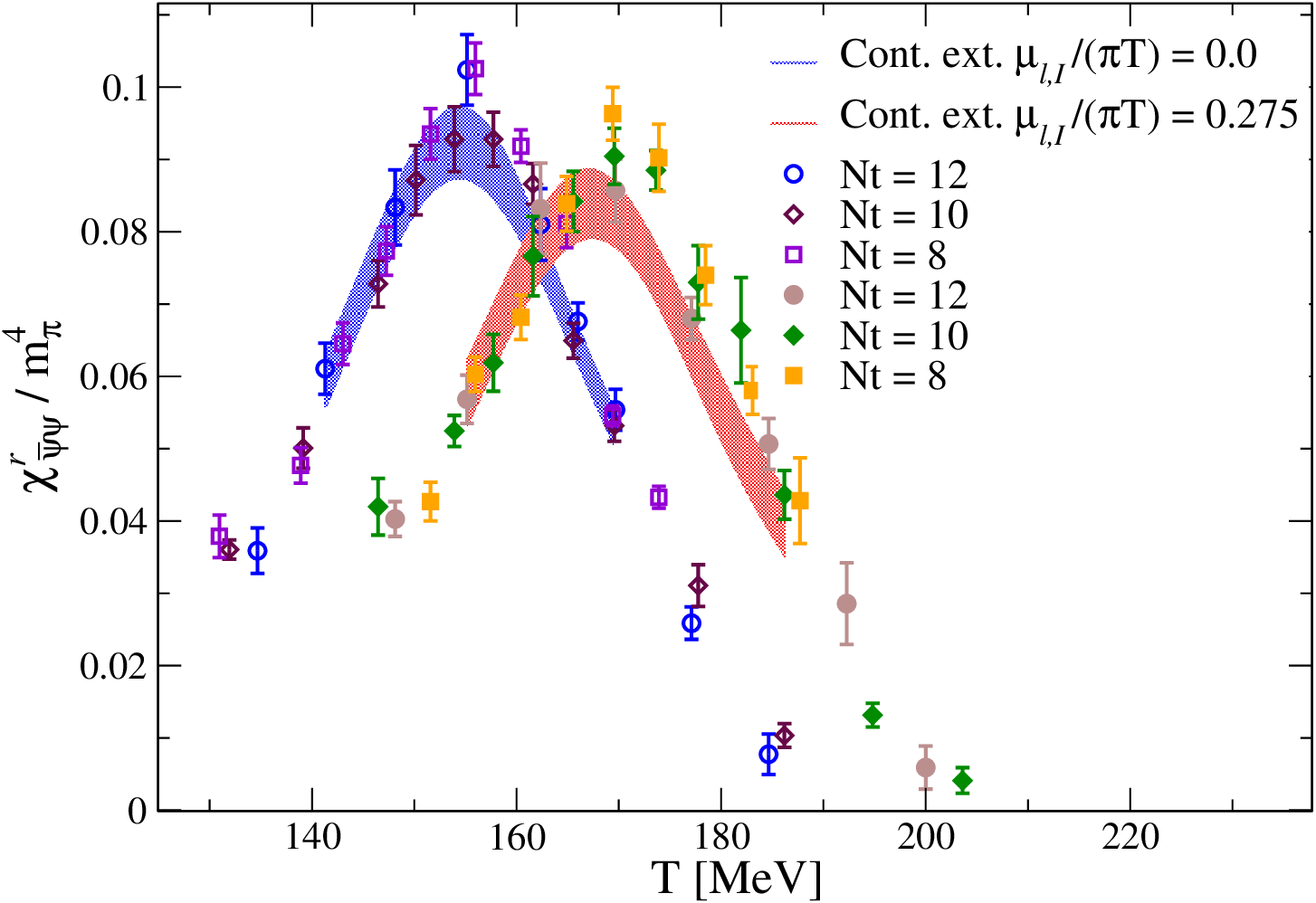} 
\hspace{0.5cm}
\raisebox{-0.15cm}{\includegraphics[width=0.413\textwidth,clip]{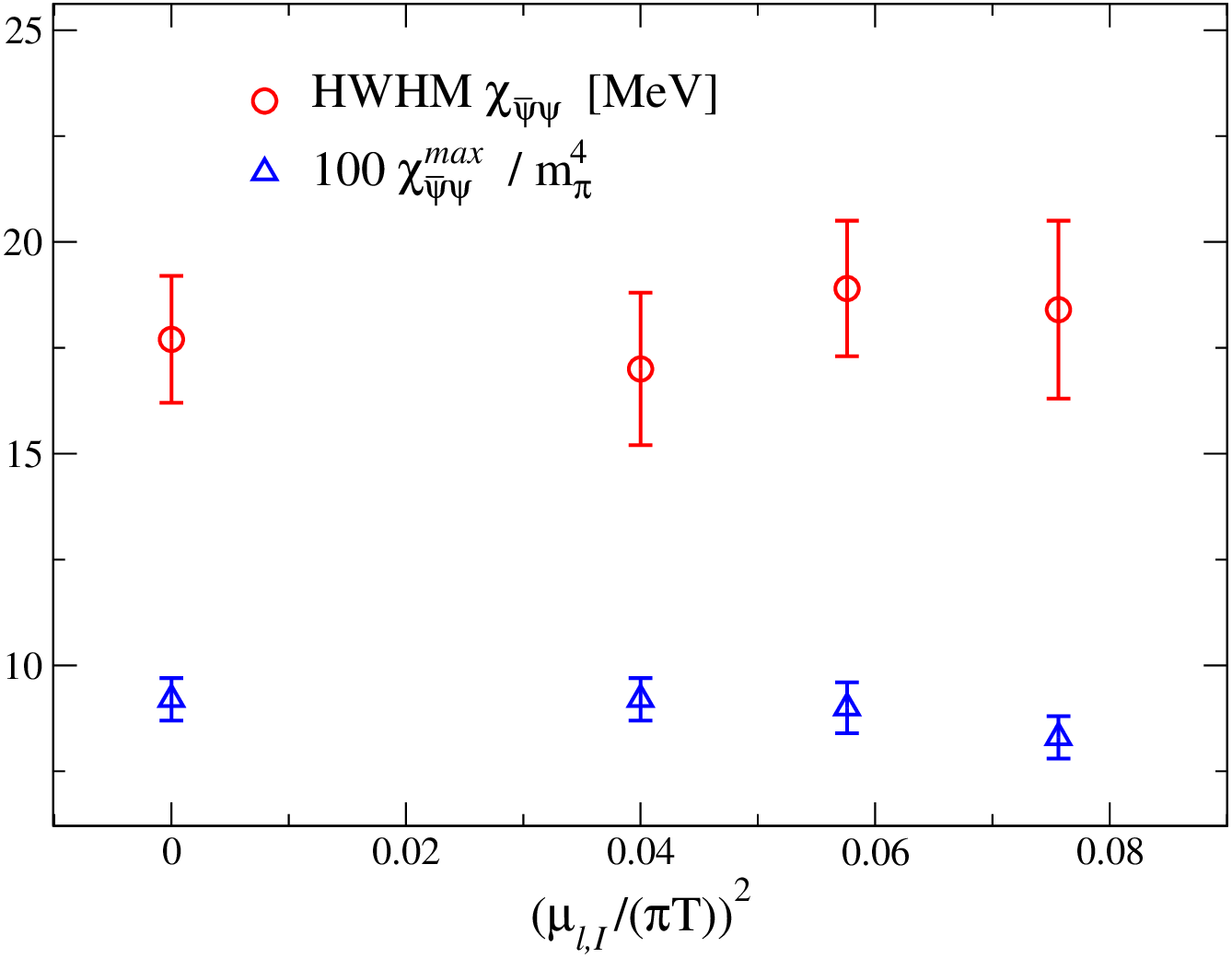}}
\caption{(left) Continuum extrapolated chiral susceptibility for two values of
the light chemical potential. (right) Dependence of the half-width at
half-maximum (HWHM) of $\chi_{\bar{\psi}\psi}$ and of the maximum value of
$\chi_{\bar{\psi}\psi}$ on the light chemical potential.}
\label{fig:crit}
\end{figure}

As a byproduct of the determination of the curvature $\kappa$, the continuum
extrapolated chiral susceptibility was obtained in \cite{Pisa} for several
values of the imaginary chemical potential, see Fig.~\ref{fig:crit} (left). The
dependence on the chemical potential of $\chi_{\bar{\psi}\psi}$ can give some
hints on the location of a critical endpoint: a critical endpoint for $\mu_B>0$
would suggest the maximum of $\chi_{\bar{\psi}\psi}$ to decrease and its
half-width at half-maximum (HWHM) to increase as the imaginary chemical
potential is increased.  None of these behaviours was however observed in the
numerical data, see Fig.~\ref{fig:crit} (right). While this is obviously not
incompatible with the existence of a critical endpoint at $\mu_B>0$, it is an
indication that (if it exists) it can not be too close to the real axis.

\end{document}